# A multiphase constitutive modeling framework for unsaturated soil behavior

G.D. Nguyen
*School of Civil, Environmental and Mining Engineering, University of Adelaide*

Y. Gan
*School of Civil Engineering, University of Sydney*

ABSTRACT: We develop a framework for constitutive modeling of unsaturated soils that has the embedded elements of lower scale grain to grain contacts. Continuum models developed from this framework will possess two different phases idealizing the solid grains and their interactions. As a consequence, two different constitutive relationships, corresponding to the grain to grain contact and bulk behavior, co-exist in a constitutive model and govern the response of the model. To be specific, grain to grain sliding under dry or wet condition is idealized and appears as a simple contact law embedded in a continuum framework. There is no need to define plastic strain, as this quantity naturally emerges at the continuum scale as the consequence of frictional sliding at the lower scale. In addition, the effective stress can be naturally worked out from the grain to grain contact law embedded in the model without being subjected to any interpretation. This, in our opinion, is a closer representation of unsaturated soil behavior, compared to existing continuum approaches that map everything onto a single stress-strain relationship. In this paper, the framework is presented in its simplest form that takes into account sliding on a single orientation. Grain to grain contact law with capillary effects is used for the demonstration of the concept, and the technical details behind it. Generalization of the framework for better representation of unsaturated soil behavior will also be sketched out.

## 1 INTRODUCTION

In unsaturated porous media, it is well known that the degree of saturation affects not only the shear strength of the material, but also its volume and hydraulic properties (Sheng 2011). For dry porous media, the repulsive force between the contacting grains can usually be accurately described by an elasto-plastic contact model (Mitchell & Soga 2005). For water-bonded particles, however, a specific attractive force exists: the meniscus produces a suction, which, in turn, generates inter-particle compressive forces, named capillary forces, whose mechanics depends on the degree of saturation of the medium. At low saturations (the pendular regime; Bear 1972), the capillary force depends on the separation between the two grains, the radius of the liquid bridge, interfacial tension and contact angle (Lian et al. 1993). At high saturations (the funicular regime; Bear 1972), negative water pressure acts all around the particle. Generally, the capillary force depends on the grain sizes and the local geometry of the grains (Scholtès et al. 2009).

The richness of the underlying micromechanics of unsaturated media response requires not only extensive and reliable micromechanical studies, but also the development of a continuum theory to accommodate the observed micromechanical details. In this sense, most (if not all) results from previous micromechanical studies are mapped to the classical framework of continuum mechanics theory. We note that on the basis of classical continuum mechanics, there is a lack of consensus on the definition of effective stress in unsaturated media (Nuth & Laloui 2008). Widely-used empirical measurements of soil-water characteristic curves (SWCCs) offer practical descriptions of a key constitutive relationship but provide limited physical understanding of these relationships and therefore limited theoretical guidance for their interpretation. In addition, existing continuum descriptions cease to be valid once softening and localized deformations occur due to the loss of homogeneity of the volume element in which stresses, strains and other macroscopic quantities are defined. Continuum assumption is no longer valid in such cases and attempts to capture the behavior of unsaturated soils exhibiting localized deformations will at best lead to phenomenological descriptions. Therefore it is clear that the enrichment of continuum mechanics framework is not only favored but also necessary because it offers a great opportunity to develop constitutive models of porous media over a wide range of saturations and loading paths.

Despite the many contributions to the micromechanical enrichments of constitutive models for un-saturated soils (e.g. Loret and Khalili 2000&2002; Pereira et al. 2005; Gray and Schrefler 2001; Nikooee et al. 2013; Borja & Koliji 2009), we wish to explore an alternative path in this paper. The key idea is to (i) observe the micromechanics, (ii) propose an abstract representation of the observed micromechanisms, and then (iii) embed them in a constitutive model. While (i) for unsaturated soil behavior is rather clear in the literature, (ii) and (iii) is less so and they are the focus here. Our work on a multi-scale continuum theory by (Nguyen et al. 2012 & 2014; Nguyen, 2013) will be used to convey the ideas associated with (ii) and (iii). Section 2 will present key characteristics of that framework to form a basis for the developments on unsaturated behavior. Simple numerical examples will be presented to demonstrate the potentials of the new approach.

## 2 A MULTI PHASE FRAMEWORK

### 2.1 *Idealization of the physics*

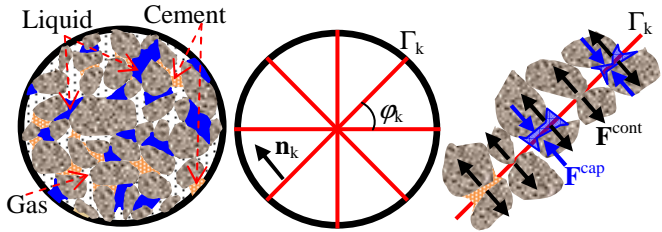

Figure 1. An RVE of unsaturated porous rocks, its multi-phase idealization, and a contact phase.

Figure 1 illustrates a volume of cemented soil under unsaturated condition. While mapping everything onto the classical continuum mechanics framework using a single stress/strain tensor has been a practically acceptable approach for many years, despite the existing limitations discussed in the introduction, we wish to explore a new path here. In Figure 1 above, the bulk behavior and contact phases are separated. While features such as porosity reduction can be lumped into the bulk behavior, grain to grain contact in our opinion can be better represented by a separated constitutive law embedded in the general continuum framework. This additional constitutive relationship is again idealized as a special cohesive-frictional law taking into account the contributions from different components of the grain to grain contacts. In this sense, the volumetric behavior is associated with the bulk, while its shear strength is governed by collective responses of several possible surfaces across which a cohesive type law is defined. The coupling between these two phases, bulk and surface, is taken into account via an equilibrium condition, in terms of traction continuity across the surface. Loosely speaking, in the above idealization, the shear response of the bulk material is weakened by several contact surfaces, each of which is inclined at an angle $\varphi$ to the horizontal (Fig. 1) and possesses an internal degree of freedom. The added complexity compared with traditional continuum representation allows for an embedded grain to grain contact law that can take into account cemented behavior, and/or bare frictional contact under dry or wet condition.

### 2.2 *Mathematical descriptions*

The mathematical descriptions closely follow the above physical idealization. For simplicity we take at first the case of a continuum weakened by a single contact surface possessing normal vector **n** (Fig. 1). The generalization of the concepts and descriptions for multi contact surfaces will directly follow. As a reasonable assumption rooting from the lumping of all inelastic behavior on surfaces, the bulk material in this case is considered elastic. This is for the sake of simplicity in the representation, as generally inelastic behavior can also be taken into account.

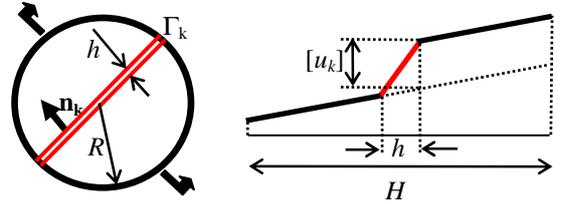

Figure 2. Bulk material crossed by a discontinuity and displacement profile across the discontinuity.

The basis roots from our previous work on a multi-scale continuum framework in which behaviors at two scales are connected (Nguyen et al. 2012 & 2014). We denote $h$ the thickness of a thin layer usually idealized as a surface of kinematic discontinuity (Fig. 2), $H$ the effective size of the RVE defined as $H=V_{RVE}/A$ with $V_{RVE}$ being the volume of the RVE and $A$ the area of the surface $\Gamma$. For a spherical RVE, $H=4R_{RVE}/3$, while it slightly varies with the orientation for a cubic RVE (see Nguyen et al. (2012) for details). We however can leave the issues of RVE shape out at the moment when dealing with a single orientation. For this, we have $f=h/H$ as the volume fraction of this thin layer (Nguyen et al. 2012). The composite nature of the continuum results in the following relationship between strain rates:

$$\dot{\boldsymbol{\varepsilon}} = f\dot{\boldsymbol{\varepsilon}}_k + (1-f)\dot{\boldsymbol{\varepsilon}}_o \qquad (1)$$

In the above equation, $\boldsymbol{\varepsilon}$ is the macroscopic strain, $\boldsymbol{\varepsilon}_k$ the strain inside the thin layer, and $\boldsymbol{\varepsilon}_o$ the strain in the bulk continuum. For $h<<H$, e.g. sliding is localized on a plane or a very thin layer, the strain rate inside this layer can be approximated as:

$$\dot{\boldsymbol{\varepsilon}}_k = \tfrac{1}{h}\left(\mathbf{n}_k \otimes [\dot{\mathbf{u}}_k]\right)^s = \tfrac{1}{2h}\left(\mathbf{n}_k \otimes [\dot{\mathbf{u}}_k] + [\dot{\mathbf{u}}_k] \otimes \mathbf{n}_k\right) \quad (2)$$

where $[\dot{\mathbf{u}}_k]$ is relative velocity between opposite sides of the thin band $k$ where sliding takes place. Due to this sliding, the stress rate in the bulk continuum is relaxed as (Nguyen et al. 2012):

$$\dot{\boldsymbol{\sigma}} = \mathbf{a}_o : \dot{\boldsymbol{\varepsilon}}_o = \tfrac{1}{1-f}\mathbf{a}_o : \left(\dot{\boldsymbol{\varepsilon}} - f\dot{\boldsymbol{\varepsilon}}_k\right) \qquad (3)$$

For elastic behavior in the bulk, $\mathbf{a}_o$ denotes the elastic stiffness tensor. On the other hand, inelastic response is lumped onto the thin band and governed by the following generic constitutive relationship:

$$\dot{\boldsymbol{\sigma}}_k = \mathbf{a}_k^T : \dot{\boldsymbol{\varepsilon}}_k \qquad (4)$$

with $\mathbf{a}_k^T$ being the tangent stiffness of the material, and $\boldsymbol{\sigma}_k$ the stress inside the localization band $k$.

As can be seen we treat the material as a composite one consisting of two different phases with corresponding behaviors. These behaviors are connected via an internal equilibrium condition to maintain the continuity of traction across the boundary of the thin localization band:

$$\left(\dot{\boldsymbol{\sigma}} - \dot{\boldsymbol{\sigma}}_k\right)\cdot \mathbf{n}_k = 0 \qquad (5)$$

While our previous work (Nguyen et al. 2012 & 2014) has presented the general case with the width $h$ of the thin localization band considered finite or infinitesimal, the focus here is the latter, as sliding between grains in soils can be reasonably assumed to take place across an imaginary surface of zero thickness. In such a case of $f \to 0$, from (2) and (3) the stress rate in the bulk material becomes:

$$\dot{\boldsymbol{\sigma}} = \mathbf{a}_o : \left[\dot{\boldsymbol{\varepsilon}} - \tfrac{1}{H}\left(\mathbf{n}_k \otimes [\dot{\mathbf{u}}_k]\right)^s\right] \qquad (6)$$

From (2) and (4), the constitutive relationship across the surface of discontinuity can be transformed as followed:

$$\dot{\mathbf{t}}_k = \mathbf{a}_k^T : \tfrac{1}{h}\left(\mathbf{n}_k \otimes [\dot{\mathbf{u}}_k]\right)^s \cdot \mathbf{n}_k = \tfrac{1}{h}\left(\mathbf{n}_k \cdot \mathbf{a}_k^T \cdot \mathbf{n}_k\right)\cdot [\dot{\mathbf{u}}_k] \qquad (7)$$

which can be better expressed in a discrete form as:

$$\dot{\mathbf{t}}_k = \mathbf{K}_k^T \cdot [\dot{\mathbf{u}}_k], \text{ where } \mathbf{K}_k^T = \tfrac{1}{h}\left(\mathbf{n}_k \cdot \mathbf{a}_k^T \cdot \mathbf{n}_k\right) \qquad (8)$$

Details and physical reasoning behind the two cases of $h$ finite and $h \to 0$ can be found in Nguyen et al. (2012 & 2014). In the context of unsaturated soil behavior, the tangent stiffness $\mathbf{K}_k^T$ in (7) dictates the behavior across the surface of discontinuity, as contact laws under wet and dry conditions can be prescribed directly. These contact laws will be described in the next section.

The generalization of the above constitutive relationships to take into account sliding on multiple surfaces leads to the following set of equations:

Stress-strain: $\dot{\boldsymbol{\sigma}} = \mathbf{a}_o : \left[\dot{\boldsymbol{\varepsilon}} - \tfrac{1}{H}\int_0^\pi \left(\mathbf{n} \otimes [\dot{\mathbf{u}}(\varphi)]\right)^s d\varphi\right] \qquad (9)$

Surface contact law: $\dot{\mathbf{t}}(\theta) = \mathbf{K}_\theta^T \cdot [\dot{\mathbf{u}}(\varphi)] \qquad (10)$

Traction continuity: $\dot{\boldsymbol{\sigma}} \cdot \mathbf{n}(\theta) = \dot{\mathbf{t}}(\varphi) \qquad (11)$

## 2.3 Computational procedure

While numerical solutions to the system (9-11) together with corresponding algorithms are being investigated (Nguyen 2013), the computational procedure for the simplest case involving a single surface of discontinuity is presented here. As can be seen, the additional constitutive relationship for the contact phase is accompanied with the condition on traction continuity across the surface of discontinuity. Therefore, given a macroscopic strain rate, the corresponding stress rate can be worked out as follows. Enforcing the traction continuity (5) using the constitutive equations (6) and (8) results in:

$$\mathbf{a}_o : \left[\dot{\boldsymbol{\varepsilon}} - \tfrac{1}{H}\left(\mathbf{n}_k \otimes [\dot{\mathbf{u}}_k]\right)^s\right]\cdot \mathbf{n}_k = \mathbf{K}_k^T \cdot [\dot{\mathbf{u}}_k] \qquad (12)$$

Therefore for a given macroscopic strain rate, the velocity jump $[\dot{\mathbf{u}}_k]$ at the surface of discontinuity can be worked out from the above equation as:

$$[\dot{\mathbf{u}}_k] = \mathbf{C}^{-1} \cdot (\mathbf{a}_o : \dot{\boldsymbol{\varepsilon}} \cdot \mathbf{n}_k) \qquad (13)$$

where

$$\mathbf{C} = \mathbf{K}_k^T + \tfrac{1}{H}\left(\mathbf{n}_k \cdot \mathbf{a}_o \cdot \mathbf{n}_k\right) \qquad (14)$$

Substituting (13) into (6) leads to the stress rate in the following form:

$$\dot{\boldsymbol{\sigma}} = \mathbf{a}_o : \left[\dot{\boldsymbol{\varepsilon}} - \tfrac{1}{H}\left[\mathbf{n} \otimes \left(\mathbf{C}^{-1} \cdot (\mathbf{a}_o : \dot{\boldsymbol{\varepsilon}} \cdot \mathbf{n})\right)\right]^s\right] \qquad (15)$$

As can be seen, besides the macro (bulk) behavior, the constitutive equation (15) contains an embedded lower scale behavior representing the grain to grain contact. The material stiffness is dominated by grain-grain contact once it is activated, resulting in the overall stiffness governing by the size and behavior at different scales. Similar observations appeared in Ortega et al. (2007). Although not specifically addressed here, a length scale is involved in the constitutive response, which will help alleviate pathological features of classical continuum models in dealing with localized failure. Further details on the length scale, localized failure and implementation algorithms can be found at length in our previous work (Nguyen et al. 2012 & 2014).

## 3 GRAIN TO GRAIN CONTACT LAW

As a way to simplify the initial formulation, it is assumed that the fluid (in fact the saturation $S_r$) is uniformly distributed in the RVE over the whole course of deformation. This is not a compulsory assumption, as flow between orientations induced by localized deformation can be accounted for via an idealization of the flow between menisci. In addition, the grain to grain contact surface is at first assumed to align with the discontinuity surface, resulting in

$\delta_n$=[**u**].**n** as the gap between grains (Fig. 3). The total contact force **F** across Γ has solid (**F**$^{\text{cont}}$) and capillary (**F**$^{\text{cap}}$) components:

$$\mathbf{F} = \mathbf{F}^{\text{cap}} + \mathbf{F}^{\text{cont}} = \sum_i (\mathbf{f}_i^{\text{cap}} + \mathbf{f}_i^{\text{cont}}) = \sum_i \mathbf{f}_i \quad (16)$$

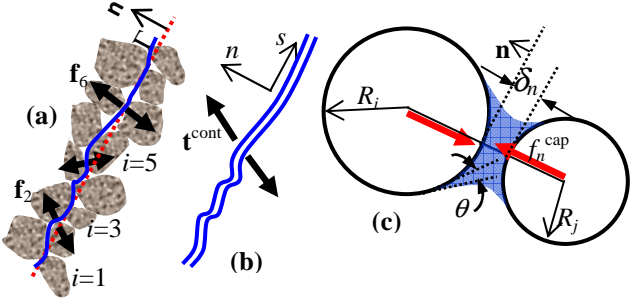

Figure 3. (a) grain contacts across an imaginary surface Γ; (b) solid frictional contact; (c) capillary force for an idealized contact pair consisting of circular grains.

From (16), the stress (or traction) across the surface of discontinuity is:

$$\mathbf{t} = \frac{\mathbf{F}}{A} = \frac{1}{A}\sum_i \mathbf{f}_i = \frac{1}{A}\sum_i \left(\mathbf{f}_i^{\text{cap}} + \mathbf{f}_i^{\text{cont}}\right) = \mathbf{t}^{\text{cap}} + \mathbf{t}^{\text{cont}} \quad (17)$$

where $A$ is the area of the discontinuity surface and $i$ denotes a contact pair across a discontinuity with grain-grain contact force $\mathbf{f}_i$. As can be seen, the macroscopic stress **t** is defined over the surface from microscopic components of the contact law. The grain-grain contact force $\mathbf{f}_i$ depends on various factors including the grain sizes, gap, and liquid bridge properties. The traction **t** on the surface Γ is expressed in general form as a statistically homogenized one based on the grain size distribution (gsd) $p(D)$:

$$\mathbf{t} = \frac{\alpha N_c}{A}\int_{D_{\min}}^{D_{\max}}\int_{D_{\min}}^{D_{\max}} p(D_1)p(D_2)\mathbf{f}(D_1,D_2)\,dD_1 dD_2 \quad (18)$$

For simplicity, the probability of one contact between grains having sizes of $D_1$ and $D_2$ is assumed to be $p(D_1)p(D_2)$. In Eq. (18), $D_{\max}$ and $D_{\min}$ are maximum and minimum grain sizes, respectively, $N_c$ the coordination number of the contact, and $\alpha$ a scaling factor proportional to the ratio between the grain size and RVE size ($H$ is the effective size of the RVE):

$$\alpha \sim A/D_{\text{mean}}^2 \approx (V_{RVE}/H)^2/D_{\text{mean}}^2 \quad (19)$$

At this point, we make assumptions to simplify the current formulation for the implementation in our existing numerical code, while leaving space for future developments. We assume all grain contacts across the plane Γ (Fig. 3) are in pairs and a pair of 2 contacting grains has the same grain size (e.g. $R_i=R_j$ in Fig. 3c). In addition, if the solid part of the contact can be separated and lumped into a simple Mohr-Coulomb contact with friction coefficient $\mu$ on the surface Γ, the homogenization in (18) can be written for only the capillary force. Therefore we have:

$$\mathbf{t}^{\text{cap}} = \frac{\alpha N_c}{A}\int_{D_{\min}}^{D_{\max}} p(D)\mathbf{f}^{\text{cap}}(D)\,dD \quad (20)$$

The stiffness $\mathbf{K}_k^T$ needed for the computation can then be derived as follows:

$$\mathbf{K}_k^T = \frac{\partial \mathbf{t}}{\partial [\mathbf{u}]} = \frac{\partial \mathbf{t}^{\text{cont}}}{\partial [\mathbf{u}]} + \frac{\alpha N_c}{A}\int_{D_{\min}}^{D_{\max}} p(D)\frac{\partial \mathbf{f}^{\text{cap}}}{\partial [\mathbf{u}]}\,dD \quad (21)$$

The first term in the above expression is the solid contact, which takes linear contact law with friction. The governing equations are written in local coordinate system (Fig. 3b) as an elasto-plastic problem.

$$t_n^{\text{cont}} = H(-u_n)K_n u_n\,;\ t_s^{\text{cont}} = K_s\left(u_s - u_s^f\right) \quad (22)$$

$$y = \left|t_s^{\text{cont}}\right| - \mu\left\langle -t_n^{\text{cont}}\right\rangle = 0 \quad (23)$$

In the above equations, $K_n$ and $K_s$ are contact stiffnesses which are theoretically infinity for the case $h\to 0$ in this study (Nguyen et al. 2014); $u_s^f$ denotes the relative sliding between opposite faces of the discontinuity Γ, and <> the Macaulay function. The Heaviside function $H(x)$ is introduced in (22) to take into account the unilateral contact, while frictional behavior is governed by the friction coefficient $\mu$.

The second term in (21) is the grain-grain capillary interaction $\mathbf{f}^{\text{cap}}$ which is the solution of Laplace equation (Soulie et al. 2006). Closed form exact solutions do not exist and a range of approximated solutions are given in the literature (e.g. Lian et al. 1993; Soulie et al. 2006; and Richefeu et al. 2006). They have been used extensively for DEM simulations. Within this framework, any capillary interaction laws can be used and we do not go into a debate on which one is the best. In this sense, the law proposed by Soulie et al. (2006) is employed, as it is simple while giving satisfactory performance. In Fig. 3c, the capillary force $\mathbf{f}^{\text{cap}}$ between two grains of radius $R_i$ and $R_j$ is normal to the contact plane and its normal component given by (Soulie et al. 2006):

$$f_n^{\text{cap}} = \pi\gamma_s\sqrt{R_iR_j}\left[c+e^{a\delta_n/R_i+b}\right] \quad (24)$$

where $\gamma_s$ is liquid surface tension. Coefficients $a$, $b$, and $c$ are expressed in terms of the liquid bond volume $V_b$, particle-liquid-gas contact angle $\theta$, and the bigger radius $R_i$ (Fig. 3c):

$$a = -1.1\left(V_b/R_i^3\right)^{-0.53} \quad (25)$$

$$b = 0.48 - (0.148\theta^2 + 0.0082)\ln(\tfrac{V_b}{R_i^3}) - 0.96\theta^2 \quad (26)$$

$$c = 0.0018\ln\left(V_b/R_i^3\right)^{-0.53} + 0.078 \quad (27)$$

The cut off value $\delta_n^{\max}$ (corresponding to $f_n^{\text{cap}}=0$) depends on the liquid bond volume $V_b$, and particle-liquid-gas contact angle $\theta$ (Fig. 3c) (Lian et al. 1993; Soulie et al. 2006):

$$\delta_n^{\max} = \left(1 + \frac{\theta}{2}\right) V_b^{1/3} \qquad (28)$$

To link the above contact law to the proposed continuum framework, the liquid bond volume $V_b$ is related to the saturation $S_r$, which is assumed to be uniform over the whole RVE in this paper. In this sense, $S_r$ is the same across the entire grain size distribution, resulting in the variation of $V_b$ with respect to the harmonic mean grain size $R$. As stated earlier, this preliminary assumption can be removed if flow between menisci due to localization effects is taken into account. Denoting $\phi$ the porosity of the RVE, and $V_g$ the solid volume of a grain, we have:

$$V_b = \frac{2}{N_c} \frac{\phi}{1-\phi} S_r V_g \qquad (29)$$

## 4 CAPILLARY EFFECTS ON MODEL RESPONSES

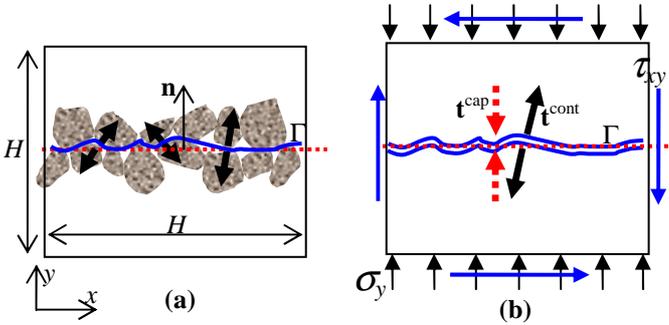

Figure 4. Effects of capillary forces: (a) grain contacts across an imaginary surface $\Gamma$; (b) idealization of the solid grain-grain and capillary contacts.

Simple numerical examples are used here to demonstrate the proposed concept. Sliding on a single orientation is considered, in which a simple Mohr-Coulomb friction law on the surface $\Gamma$ is enhanced with capillary effects (Fig. 4), the latter takes the form (24). The dilation due to surface roughness (Fig. 4b) as a representation of the true contact (Fig. 4a) can be incorporated using a multi-scale contact law across $\Gamma$. In addition, the incorporation of compaction due to grain crushing has also been sketched out and discussed (Nguyen 2013; Shen et al. 2013). They both are planned for our future work.

The bulk behavior is elastic with Young modulus $E=30$MPa and Poisson's ratio $\nu=0.2$. For the solid phase of the contact, the friction coefficient is taken as $\mu=0.35$, and contact stiffnesses $K_n$ and $K_s$ determined based on Eq. (8), using $h=H/10^6$. For the capillary phase, we assume mono-disperse sample with grain size $R=0.1$mm. The corresponding coordination number is assumed as $N_c \approx 6$. The gsd becomes a Dirac delta function and Eq. (21) reduces to:

$$\mathbf{K}_k^T = \frac{\partial \mathbf{t}}{\partial [\mathbf{u}]} = \frac{\partial \mathbf{t}^{cont}}{\partial [\mathbf{u}]} + \frac{\alpha N_c}{A} \frac{\partial \mathbf{f}^{cap}}{\partial [\mathbf{u}]} \qquad (30)$$

For the example, the RVE size is $H=10$mm (Fig. 4). Therefore under plane strain condition, the area of the contact is $A=1*H=10$mm$^2$. Other parameters are as follows: porosity $\phi=0.2$, particle-liquid-gas contact angle $\theta=30^o$, liquid surface tension $\chi_s=7.3*10^{-5}$ N/mm. For the above parameters, the effect of saturation on capillary law is depicted in Fig. 5.

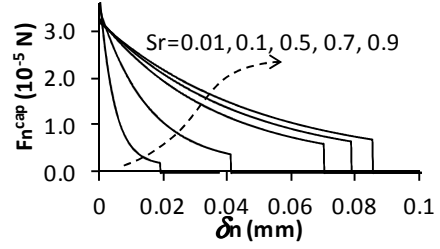

Figure 5. Effect of saturation $S_r$ on capillary law.

### 4.1 Tensile strength

The tensile behavior of the model, with cohesive strength due to capillary suction, is explored by varying the saturation $S_r$ (1%-90%), and Young modulus $E$ (around its given value of 30MPa).

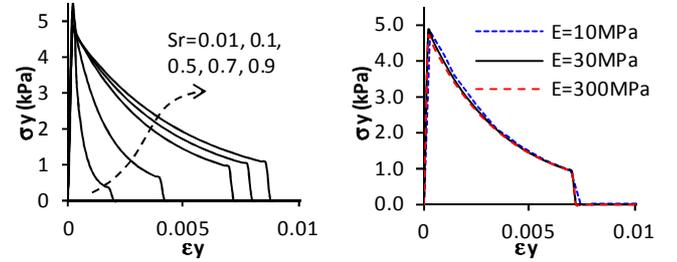

Figure 6. Effect of saturation $S_r$ (a, left) and bulk stiffness $E$ (b, right; using $S_r$=0.5) on tensile response.

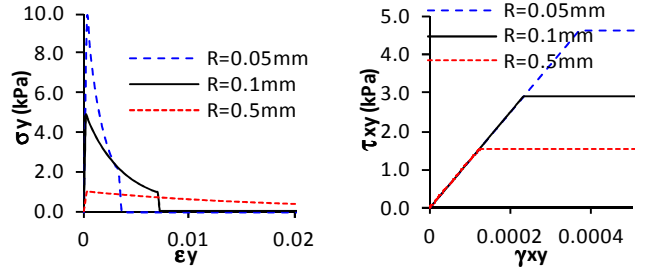

Figure 7. Effect of grain size on tensile (a, left) and shear (b, right) responses (using $S_r$=0.5).

Figure 6 shows the effects of $S_r$ and $E$ on the tensile behavior of the material. The trend in the response is similar to the capillary law depicted in Figure 5, as the liquid bond behavior is dominant in this case with tensile strain localized onto the capillary contact, due to the much higher stiffness of the bulk. This is automatically captured by the constitutive structure of our model (see Eqs. 14-15). Figure 7a shows the effect of grain size on cohesive strength of the material. Since the cohesive strength scales with $1/R^2$ (Eq. 19) while capillary force scales with $R$ (Eq. 24), the material becomes stronger in tension when the grain size reduces. However the strain at rupture decreases for decreasing grain size.

### 4.2 Shear strength

In consistence with the observation in tension, the shear strength of the material increases with decreas-ing grain size (Fig. 7b). The shear response under zero pressure (normal stress $\sigma_y$ in this case) is depicted in Figure 8a and compared with the shear strength under a pressure of 3.33kPa.

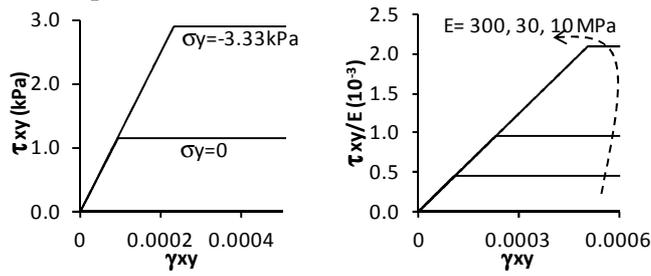

Figure 8. Effect of pressure (a, left) and bulk stiffness $E$ (b, right) on shear strength (using $S_r$=0.5).

The material model in this case possesses shear strength even under zero pressure, thanks to the capillary effects. The saturation $S_r$ has little effects on the shear strength in this case, due to little change in $\delta_n$ during compression. On the other hand, the capillary effects, via the non-dimensional number $\tau_{xy}/E$, are getting stronger with decreasing stiffness (Fig. 8b). In that figure, the same vertical strain $\varepsilon_y$=-10$^{-4}$ was applied before shearing started. While the capillary contribution remains unchanged, the solid contact forces becomes weaker with decreasing stiffness, resulting in greater capillary effects for softer materials.

## 5 CONCLUSIONS

A new approach was proposed to embed micromechanical components of the grain-grain contact in constitutive models. It utilizes the multiphase nature of soil behavior in the simplest form and provides room to accommodate micromechanical characteristics that are well known in unsaturated soil behavior. The soil response is decomposed into localized and bulk phases to reflect the grain to grain contact and bulk behavior, respectively. In this sense, any kind of contact at the grain scale can be employed to model the response of the material. We demonstrated essential features of the new constitutive modeling framework and scope for future developments. It should be emphasized that the proposed framework is just an idealization of the much richer grain scale phenomena and aims to reduce the number of ad hoc or empirical laws. The numerical results at this stage just show the trends rather than the accurate predictions. However the promising features of the new approach are highlighted and its current limitations addressed for improvements in the future.

## 6 ACKNOWLEDGEMENTS

Support from the Australian Research Council to G.D. Nguyen (projects DP110102645 and DP140100945) and Y. Gan (project DE130101639) is gratefully acknowledged.